\documentstyle[epsfig,osa,manuscript]{revtex}

\begin{document}

\title{Surface Quasigeostrophic Turbulence : The Study of an Active Scalar}
\author{Jai Sukhatme\footnote{email : jai@geosci.uchicago.edu} and Raymond T. Pierrehumbert
\footnote{email : rtp1@midway.uchicago.edu}}
\address{Dept. of Geophysical Sciences, University of Chicago, Chicago
, IL 60637}
\date{\today}
\maketitle

\begin{abstract} 

We study the statistical and geometrical properties of the potential temperature (PT)
field in the Surface Quasigeostrophic (SQG) system of equations.
In addition to extracting
information in a global sense via tools such as the power spectrum, the g-beta 
spectrum and
the structure functions we explore the local nature of the PT field by means of the
wavelet transform method. The primary indication is that an initially smooth
PT field becomes rough
(within specified scales), though in a qualitatively sparse fashion.
Similarly, initially 1D iso-PT contours (i.e., PT level sets) are seen to acquire a fractal
nature. Moreover, the dimensions of the iso-PT contours satisfy existing
analytical bounds.
The expectation that the roughness will manifest itself in the singular nature of the gradient
fields is confirmed via the multifractal nature of the dissipation field.
Following earlier work on the subject,
the singular and oscillatory nature of the gradient field is investigated by
examining the scaling of a probability measure and a sign singular measure respectively.
A physically motivated derivation of the relations between the variety of
scaling exponents is presented, the aim being to bring out some of
the underlying assumptions which seem to have gone unnoticed in
previous presentations.
Apart from concentrating on specific properties of the SQG system, a broader theme
of the paper is a comparison of the diagnostic inertial range properties of the 
SQG system with both the 2D and 3D
Euler equations.

\end{abstract}


\clearpage 

\narrowtext

\section{Introduction} 

In the Quasigeostrophic (QG) framework \cite{Pedlosky}, a simplification of the Navier
Stokes equations for describing the motion of a stratified and rapidly rotating fluid in a
3D domain, there are two classes of problems that immediately come to attention. The first
(Charney type) are the ones where attention is focussed on the interior of the domain; the
temperature is uniform along the boundaries and they play no dynamical role in the
evolution of the system. The other class (Eady type) of problems lead to the 
Surface Quasigeostrophic (SQG) equations. The potential vorticity in the 3D interior
is forced to be zero and the dynamical problem is controlled by the evolution
of the potential temperature at the 2D boundaries. Working with a single lower
boundary (assuming all fields to be well behaved as $z \rightarrow \infty$), the 
equations making up the SQG system can be expressed as \cite{Held-JFM},\cite{RTP-CSF},

\begin{eqnarray}
{\partial}_{t}\theta + u^{i} {\partial}_{i} \theta= D_{t}\theta = 0 ~, z=0 \nonumber \\
\theta = {\partial}_{z} \psi  = {(-\triangle)}^{\frac{1}{2}} \psi \nonumber \\ 
\textrm{where} ~~\nabla^2 \psi = 0, z>0 ~\textrm{and} ~~\vec{u} = {\nabla}^{\perp}\psi
\label{1}
\end{eqnarray}
Here $\theta$ is the potential temperature (it is a dynamically active scalar due to the coupling of 
$\theta$ to $\psi$), 
$\psi$ is the geostrophic streamfunction, 
$\vec{u}$ the geostrophic
velocity, ${\nabla}^{\perp} \equiv (-{\partial}_{y},{\partial}_{x})$, $\triangle$ is 
the horizontal Laplacian, $\nabla^2$ is the full 3D Laplacian, the  
operator ${(-\triangle)}^{\frac{1}{2}}$ is
defined \cite{Const-NL} in Fourier space via $\widehat{{(-\triangle)}^{\frac{1}{2}} \psi(k)} =
|k| \hat{\psi}(k)$  
and $i = x,y$. 
Recall that the 2D Euler equations (for an incompressible fluid) in vorticity form are,

\begin{eqnarray}
	{\partial}_{t}\xi + u^{i} {\partial}_{i} \xi= D_{t}\xi = 0  \nonumber \\
	\xi = \triangle \psi ~\textrm{with} ~ \vec{u} = {\nabla}^{\perp}\psi 
\label{2}
\end{eqnarray}
where $\xi$ is the vertical component vorticity. 
The similarity in the evolution equations for
$\theta$ and $\xi$ has been explored in detail by Constantin, Majda and Tabak
\cite{Const-NL},\cite{Majda-P}.  It can be seen that the structure of conserved quantities in
both equations is exactly the same. To be precise, just as $f(\xi)$,$\int \psi \xi$ are 
conserved by 
the 2D Euler equations similarly $f(\theta)$,$\int \psi \theta$  are conserved in the SQG system.
The basic difference in the above two systems is the
degree of locality of the active scalar. For the 2D Euler equations the free space Green's
function behaves as $\ln{(r)}$ implying a
$\frac{1}{r}$ behaviour for the velocity field due to point vortex at the origin. In contrast, in
the SQG equations the free space Green's function has the form $\frac{1}{r}$ implying a much
more rapidly decaying $\frac{1}{r^2}$ velocity field due to a point "PT vortex" at the origin
\cite{Held-JFM}. Or, in Fourier space
one has $\hat{\xi} = |k|^{2}\hat{\psi}$ and $\hat{\theta} = |k|\hat{\psi}$ for the 2D Euler
and SQG equations respectively \cite{Majda-P}. Hence the nature of interactions is much more
local in the SQG case as compared to the 2D Euler equations. \\

Studying the properties
of active scalars with different degrees of locality would be an interesting question in its
own right \cite{Sch} but the specific interest in the SQG equations comes from an analogy with the 3D Euler
equations. This can be seen by a comparison with the 3D Euler equations, which in vorticity
form read,

\begin{equation}
[\frac {D{\omega}}{Dt}]^{j} = \partial_{i} v^{j} {\omega}^{i}
\label{4}
\end{equation}
where $\vec{v}$ is a divergence free velocity field, $\vec{\omega} (=\nabla \times \vec{v})$
is the vorticity and $i,j = x,y,z$.
Introducing $\vec{V} = \vec{{\nabla}^{\perp}\theta}$, 
a "vorticity" like quantity for the SQG system which satisfies 
(differentiating Eq. (\ref{1}) and using incompressibilty),

\begin{equation}
[\frac {DV}{Dt}]^{j}  = \partial_{i} u^{j} V^{i}
\label{5}
\end{equation}
Identifying $\vec{V}$ in Eq. (\ref{5}) with $\vec{\omega}$ in Eq. (\ref{4}) 
it can be seen 
\cite{Const-NL} that the level sets of $\theta$ are geometrically analogous 
to the vortex lines for the 3D
Euler equations. Similar to the question of a finite time singularity in the 3D Euler
equations (which is thought to be physically linked to the stretching of vortex tubes), 
in the SQG system one
can think of a scenario where the intense stretching (and bunching together) 
of level set lines during the evolution
of a front leads to the development of shocks in finite time. The issue of treating
the SQG system as a testing ground for finite time singularities has 
generated interest \cite{Const-NL},\cite{Majda-P},\cite{Const-PRE1},\cite{Okhitani-PHF},\cite{Cor} 
in the mathematical community and the reader is referred to the aforementioned papers for 
details regarding this issue. 
\\

In view of the similarities between the SQG and 2D Euler equations and 
the level set
stretching analogy with the 3D Euler system, it is natural to 
inquire into
the statistical/geometrical properties of the SQG active scalar within an
appropriately defined "inertial range". 
The broad aim is to
compare these properties with the large body of work available for the 2D and 3D
Euler equations. In the second section we examine the PT field via global (power spectrum, structure
functions, ($\beta,g(\beta)$) spectrum) and local (wavelets) methods. One of the few
rigorous estimates that exist in fluid turbulence is that for the level set dimensions. 
The extraction of these dimensions and their agreement with analytical bounds
is demonstrated.
The third section
is devoted to the examination of the dissipation field, the generalized dimensions of a
measure based upon the dissipation field are calculated and commented upon. In the fourth
section we
focus our attention on the gradient fields, a simple derivation of the relation between
the variety of scaling exponents is presented and the underlying assumptions are 
clearly stated. The failure of the cancellation exponent is demonstrated and a simple
example is presented so as to put some of the ideas in perspective. 
\\

\section{The Potential Temperature Field} 

\subsection{The Power Spectrum and the Structure Functions} 

A pseudo-spectral technique was employed to solve Eq. (\ref{1}) 
numerically on a $2048 \times 2048$ grid. Linear terms are handled exactly using
an integrating-factor method, and nonlinear terms are handled by a third-order Adams-Bashforth
scheme (fully de-aliased by the 2/3 rule method). 
The calculations were carried out for freely decaying turbulence.  The initial conditions
consisted of a large-scale random field, specifically a random-phase superposition of
sinusoids with total wavenumber approximately equal to 6, in units where the gravest mode
has unit wavenumber. 
Potential temperature variance is dissipated at small scales by $\nu\nabla^2\theta$ diffusion.
Based on the typical velocity $U$ and scale $L$ of the initial condition, one may define
a Peclet number $UL/\nu$. The calculations analyzed here were carried out for a 
Peclet number of 2500.  After a short time, the spectrum develops a distinct inertial
range. As time progresses, energy and $\theta$ variance are dissipated at small scales,
the amplitude decreases, and the effective Peclet number also decreases. After sufficient
time, the flow becomes diffusion-dominated and the inertial range is lost.  Analysis
of other cases, not presented here, indicates that the results are not sensitive to the
time slice or the Peclet number, so long as the Peclet number is sufficiently large
and the time slice is taken at a time when there is an extensive inertial range.
\\

\begin{figure}
\begin{center}
\epsfxsize=9.0 cm
\epsfysize=9.0 cm
\leavevmode\epsfbox{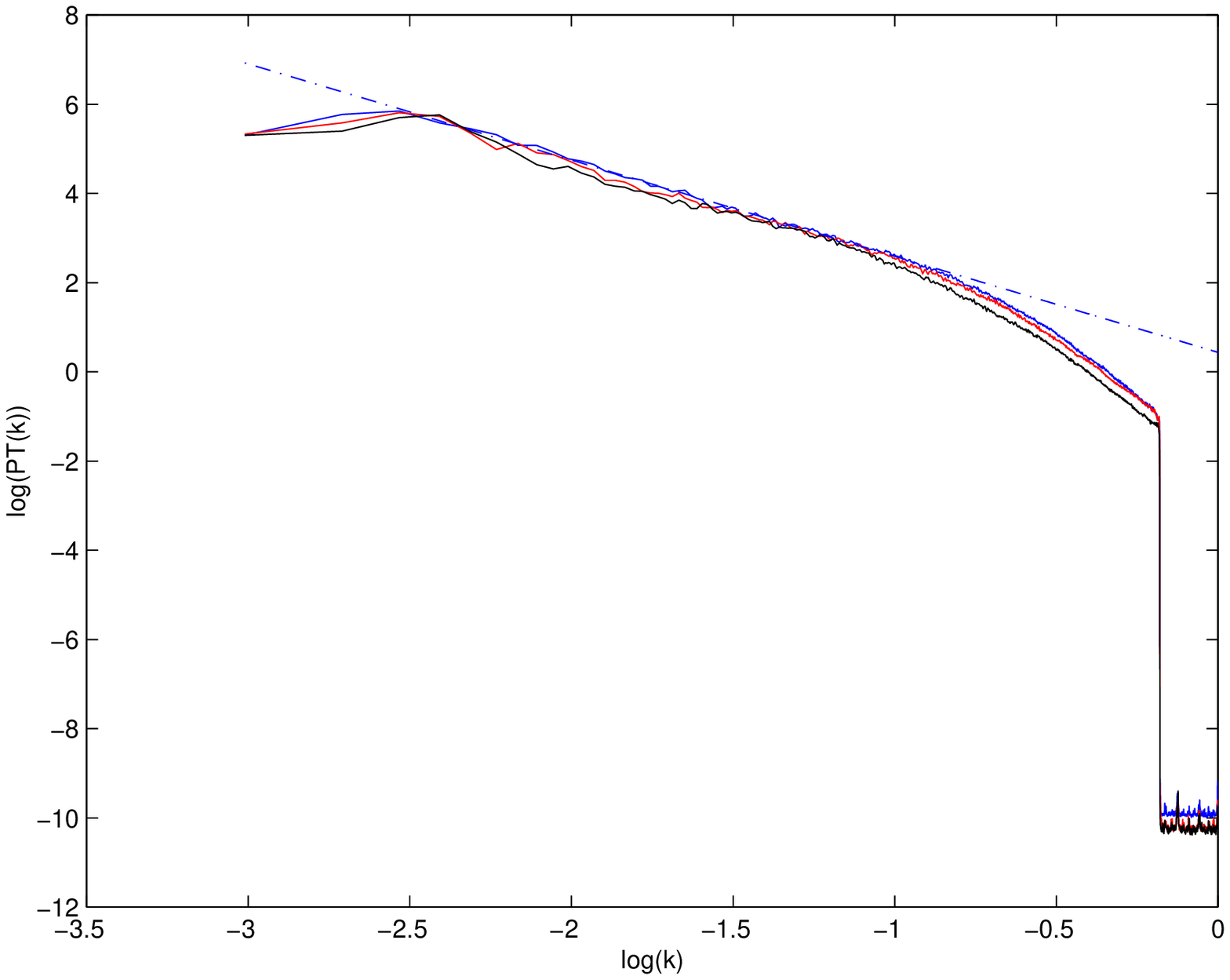}
\caption{Power spectrum of the PT field for a variety of stages.
The dashed line (extracted from the stage which has the largest inertial range)
has a slope of $-2.15$.}
\label{fig:fig1}
\end{center}
\end{figure}

\begin{figure}
\begin{center}
\epsfxsize=9.0 cm
\epsfysize=9.0 cm
\leavevmode\epsfbox{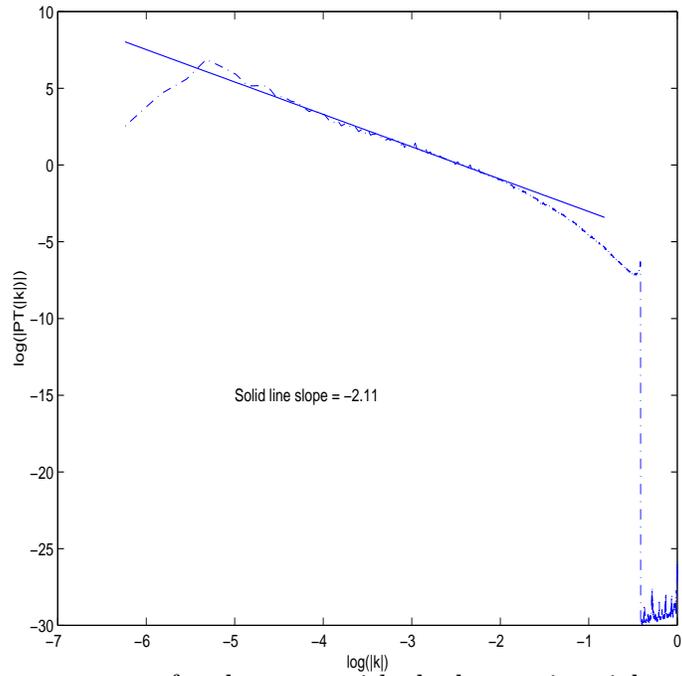}
\caption{The 2D power spectrum for the stage with the largest inertial range. The
solid line has a slope of $-2.11$.}
\label{fig:fig1c}
\end{center}
\end{figure}

\begin{figure}
\begin{center}
\epsfxsize=10.0 cm
\epsfysize=9.0 cm
\leavevmode\epsfbox{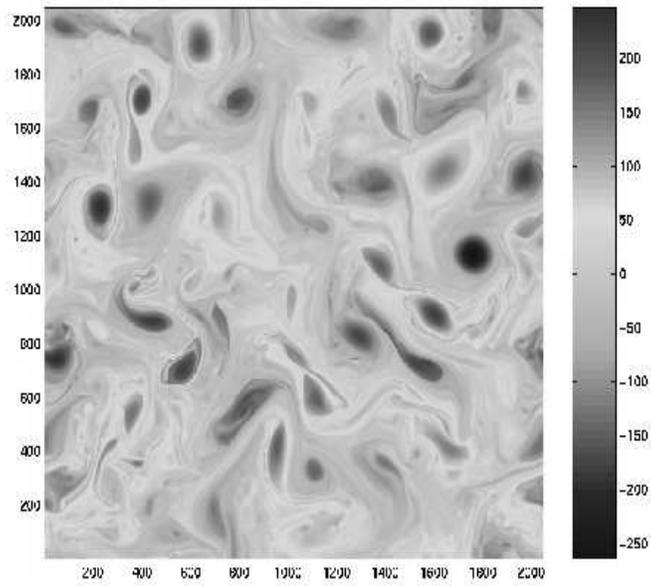}
\caption{Snapshot of the PT field which showed the largest inertial range.}
\label{fig:fig1b}
\end{center}
\end{figure}

The mean 1D power spectra
from different stages of evolution can be seen in Fig. \ref{fig:fig1}. 
As these are decaying simulations the structure in the PT field is slowly dying out.
The resulting increase in smoothness of the PT field can be seen 
via the roll off of the spectrum during the later stages.
In spite of this a fairly clean power law is visible for a sizeable "inertial range" (other runs
with large scale initial conditions posessing various amounts of energy show similar
behavior). 
We choose to concentrate on the 
particular stage which has the largest inertial range.
The 2D power spectrum for this stage can be seen in Fig. \ref{fig:fig1c} and a snapshot 
of the PT field itself can be
seen in Fig. \ref{fig:fig1b}. Interestingly the 2D power spectrum seems to roll off at
larger wavenumbers as compared to the 1D spectrum. In this stage the spectral slope
(from the 1D spectra) between the scales 256 to 8 (the scales are in terms of grid size)
is $\approx -2.15$ (the other runs also showed slopes steeper than -2). The 
slope from the 2D spectrum is $\approx -2.11$ (due to the early roll off of the 
2D spectrum, this slope is extracted between the scales 128 to 8). 
Previous decaying simulations \cite{Okhitani-PHF} obtained values near $-2$ and
seem to be consistent with our observations. A slope as steep as this suggests that 
the field being examined
is smoother than expectations from a similarity hypothesis (which yields 
a $\frac{-5}{3}$ slope \cite{RTP-CSF}). 
A closer look indicates (Fig. \ref{fig:fig1b}) that the field is
composed of a small number of "coherent structures" superposed upon a background which
has a filamentary structure consisting of very fine scales. 
This immediately brings to mind the studies on vorticity
in decaying 2D turbulence \cite{Benzi-EPL},\cite{Benzi-Math},\cite{Mizu-Jap} wherein a similar
coherent structure/background picture was found to exist. Further analysis indicated
that the vorticity field possessed normal scaling whereas a measure based upon 
the gradient of the vorticity (precisely the enstrophy dissipation) was multifractal. To
proceed in this direction we introduce the  
generalized structure functions of order $q \in \Re^+$,

\begin{equation}
         S_q(r) = <|\theta(x+r) -\theta(x)|^q>
\label{6}
\end{equation}
Here $<.>$ represents an ensemble average. The directional dependence is suppressed due 
to the assumed isotropy of the PT field. Scaling behaviour in the field implies that one
can expect the generalized structure functions to behave as \cite{Vain-PRE},

\begin{equation}
S_q(r) = C_1(q){|\triangle \theta (L_1)|}^q{(\frac{r}{L_1})}^{\zeta_q} ~;~ r_1 \le r \le L_1
\label{7}
\end{equation}
where $\zeta_q$ are the generalized scaling exponents, $C_1(q)$ is of order unity for all $q$,
$|\triangle \theta (L_1)|$ is the absolute value of the difference in $\theta$ over a scale
$L_1$. $r_1$ and $L_1$ are the inner and outer scale (8 and 256 respectively) over which 
the power law in the spectrum was observed. \\

\begin{figure}
\begin{center}
\epsfxsize=9.0 cm
\epsfysize=14.0 cm
\leavevmode\epsfbox{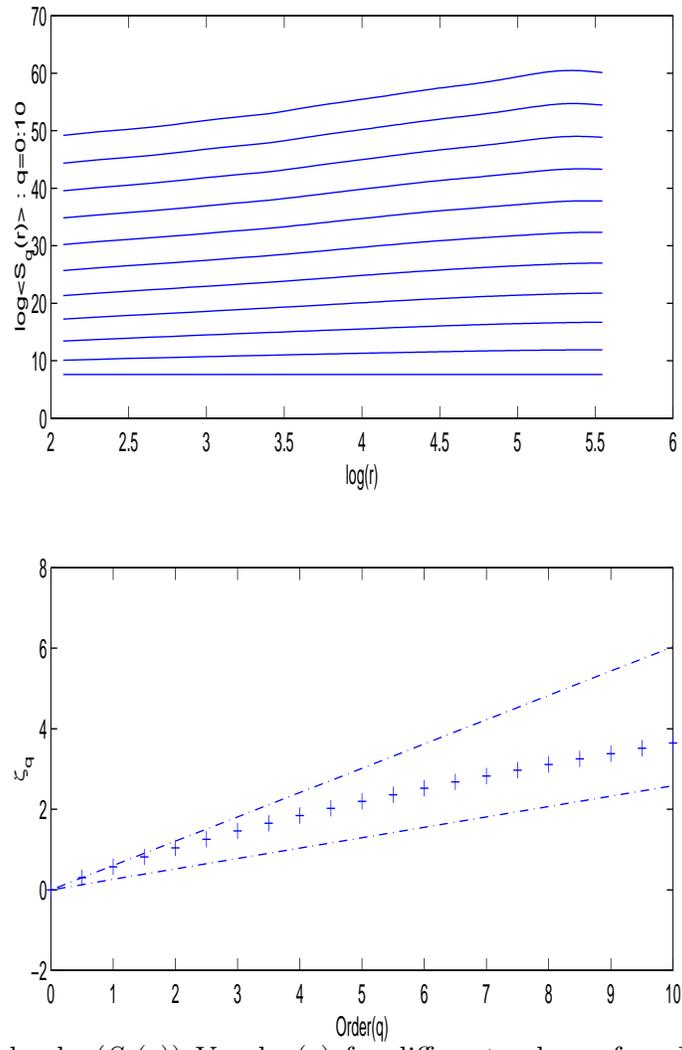}
\caption{Upper Panel : $\log(S_q(r))$ Vs. $\log(r)$ for different values of q.
Lower Panel : Scaling exponents for the PT field (+) and the linear scaling with $\beta_{min}$
and $\beta_{max}$.}
\label{fig:fig2}
\end{center}
\end{figure}

If the field
being examined is smooth at a scale $r$ then the gradient at this scale would be
finite and as a consequence $\zeta_q=q$ (due to the domination of the linear term in the
Taylor expansion about the point of interest) ie. the scaling would be trivial.
Conventionally normal scaling is a term reserved for linear $\zeta_q$ 
and any nonlinearity in $\zeta_q$ is referred to as anomalous scaling. In 2D turbulence
the velocity field is known to be smooth for all time if the initial conditions are 
smooth \cite{Rose-JPP} and hence \cite{Benzi-PRA} $\zeta_q$(velocity)$=q$. Also, as mentioned, 
from the analysis
of the vorticity field \cite{Benzi-EPL} the scaling exponents for the vorticity
structure functions were found to depend on $q$ in a linear fashion. 
Plots of $\textrm{log}(S_q(r))$ Vs $\textrm{log}(r)$  
for the PT can be seen in the upper panel of Fig. \ref{fig:fig2}.
In all cases the scaling is valid upto $r \sim 128$, using these plots we extracted 
$\zeta_q$ which are presented in the lower panel of Fig. \ref{fig:fig2}.
It is 
seen that the scaling is anomalous and in fact a best fit to the scaling 
exponents is of the form $\zeta_q = A {q}^{B}, q>0$ with $B=0.82$.
\\

For the special case of $q=2$, one can in principle relate the scaling exponent
$\zeta_2$ to the slope of the power spectrum ($n$) via \cite{Frisch-book} $n = -(1+\zeta_2)$.
This relation is only valid for $-3 < n < -1$ (note that this does not prevent the 
spectral slope from being steeper than $-3$; it just implies that $\zeta_2$ saturates at $2$ for
smooth fields and the particular relation between $\zeta_2$ and $n$ breaks down). In our
case $\zeta_2 = 1.05$ so the predicted spectral slope is $n = -2.05$ which is near
the observed mean value of $-2.15$ (or $-2.11$ from the 2D spectrum). 
Even though the scaling exponents give an idea of the roughness in the field
(anomalous scaling implying differing degrees of roughness) there is a certain
unsatisfactory aspect about the structure functions, namely, there is no
estimate of "how much" of the field is rough. The following subsection aims to address
this very issue.
\\

\subsection{The $(\beta,g(\beta))$ spectrum}

In scaling literature the roughness of a field is specified by means of 
an exponent $\beta$ ($>0$) defined as \cite{Bert-PRE},

\begin{equation}
	 |\theta(x+r) -\theta(x)| \sim |r|^{\beta(x)}
\label{8}
\end{equation}
Here $\beta$ is a function of position 
and it refers to the fact that the derivative of $\theta$ will be unbounded as $r
\rightarrow 0$ if $\beta < 1$. As mentioned previously there is a lower scale associated
with the problem  
so technically nothing is blowing up and in effect $\beta < 1$ represents the regions where
the derivative will be large as compared to the rest of the field. Note that
$\theta$ itself cannot be singular due its conserved nature. The focus is on whether an 
initially smooth $\theta$ field becomes rough so as to cause the gradient fields to 
experience a singularity. It is clear
that Eq. (\ref{8}) is by itself of not much use in characterizing $\theta$ (as $\beta$
depends upon position), in fact a global view of the specific degrees of rougness of $\theta$
can be attained via the $(\beta,g(\beta))$ spectrum which we introduce next. 
\\

\begin{figure}
\begin{center}
\epsfxsize=9.0 cm
\epsfysize=9.0 cm
\leavevmode\epsfbox{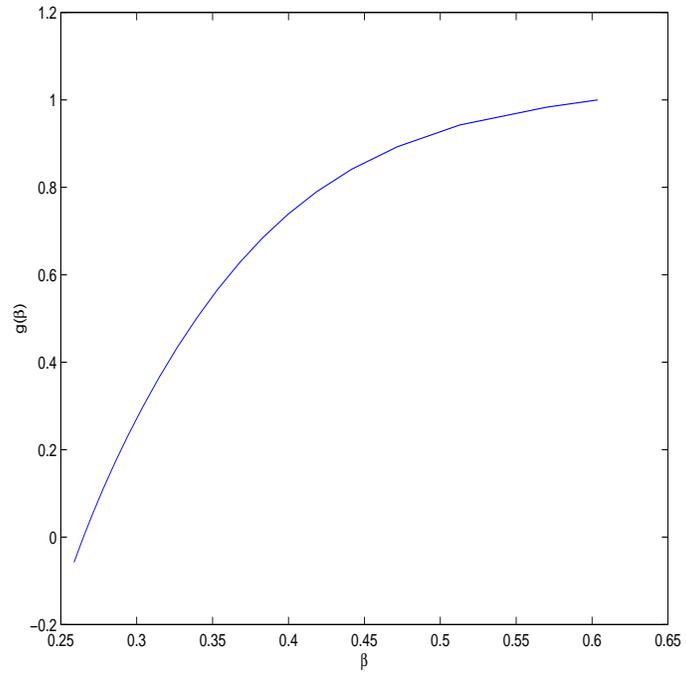}
\caption{The $(\beta,g(\beta))$ spectrum}
\label{fig:fig3}
\end{center}
\end{figure}

An iso-$\beta$ set is defined as the set of all $x$'s where $\beta(x) = \beta$ and
$g(\beta)$ is the dimension
(to be precise $g(\beta)$ should be viewed in an probabilistic fashion \cite{Frisch-book})
of an iso-$\beta$ set.
The dimensions of iso-$\beta$ sets are derived by
Frisch \cite{Frisch-Proc}. Briefly, 
at a scale $r$ the probability of encountering a particular value of
$\beta$ is proportional to $r^{d-g(\beta)}$ (where $d=2$ for the 2D PT field and
$d=1$ for 1D cuts of the PT field). 
By using a steepest descent argument in the
integral for the expectation value of $|\theta(x+r) -\theta(x)|^q$ one obtains
\cite{Frisch-Proc},\cite{Frisch-book},

\begin{equation}
    \zeta_q = \min_{\beta} [q\beta+ d-g(\beta)] ~ \textrm{or} ~ g(\beta)=\max_{q} [q\beta+ d- \zeta_q]
\label{9}
\end{equation}
Hence given $\zeta_q$ for a fixed $q=q_*$ using the first part of Eq. (\ref{9}),

\begin{equation}
	q_* = \frac{d g(\beta)}{d \beta} 
\label{9a}
\end{equation}
Denoting the value of $\beta$ for which Eq. (\ref{9a}) is satisfied by ${\beta}_*$ we
have,

\begin{equation}
        g({\beta}_*) = d + q_*{\beta}_* - \zeta_{q_*}
\label{9b}
\end{equation}
Notice that, as the structure functions involve moments with positive $q$, they only pick out
${\beta}$'s such that $\beta < 1$ and $g(\beta) < d$.
The $(\beta,g(\beta))$ spectrum seen in Fig. \ref{fig:fig3} hints at a 
hierarchy in roughness of the PT field. From the calculations we see that
$\beta \in [0.26,0.6]$. In Fig. \ref{fig:fig2} along with $\zeta_q$ we have 
plotted the lines corresponding to $\zeta_q = q\beta_{min}$ and $\zeta_q = q\beta_{max}$.
As is expected these lines straddle the actual scaling exponents. For smaller values
of $q$ the scaling exponents are close to $q\beta_{max}$ as $\beta_{max}$ has the 
largest associated dimension whereas for higher values of $q$ the scaling exponents reflect
the roughest regions and hence tend towards $q\beta_{min}$. 
Note that from 
Eq. (\ref{9a}) and Eq. (\ref{9b}) the
approximation $\zeta_q = A q^B, B<1, q>0$ implies that $\beta$ becomes large 
and $g(\beta) \rightarrow d$ as $q \rightarrow 0$. In essence the picture that emerges
is that even though $\theta$ appears to become rough (with differing degress of roughness),
in fact, it is the smooth regions that occupy most of the space.
\\

\subsection{The roughness of $\theta$ : a qualitative local view  }

All of the previous tools, the power spectrum, the $(\beta,g(\beta))$ spectrum and the
structure functions extracted information in a global sense. 
To get a picture of the actual
positions where the signal may have unbounded derivatives, and to get a qualitative
feel of the spareseness of these regions, one has 
to determine the local behaviour of the signal in
question. Recently the use of wavelets has allowed the identification of local Holder
exponents in a variety of signals. The Holder exponents are extracted by a technique
known as the wavelet transform modulus maxima (WTMM) method \cite{Muzy-PRE},\cite{Arn-PhyA}, 
\cite{Muzy-PRL},\cite{Mallat-IEEE}. The modulus maxima refers to the spatial distribution
of the local maxima (of the modulus) of the wavelet transform.
In a crude sense the previous methods used ensemble averages of the moments of differences
in $\theta(x)$ as "mathematical microscopes" whereas in the wavelet method it is the scale
parameter of the wavelet transform that performs this task. 
\\

\begin{figure}
\begin{center}
\epsfxsize=9.0 cm
\epsfysize=14.0 cm
\leavevmode\epsfbox{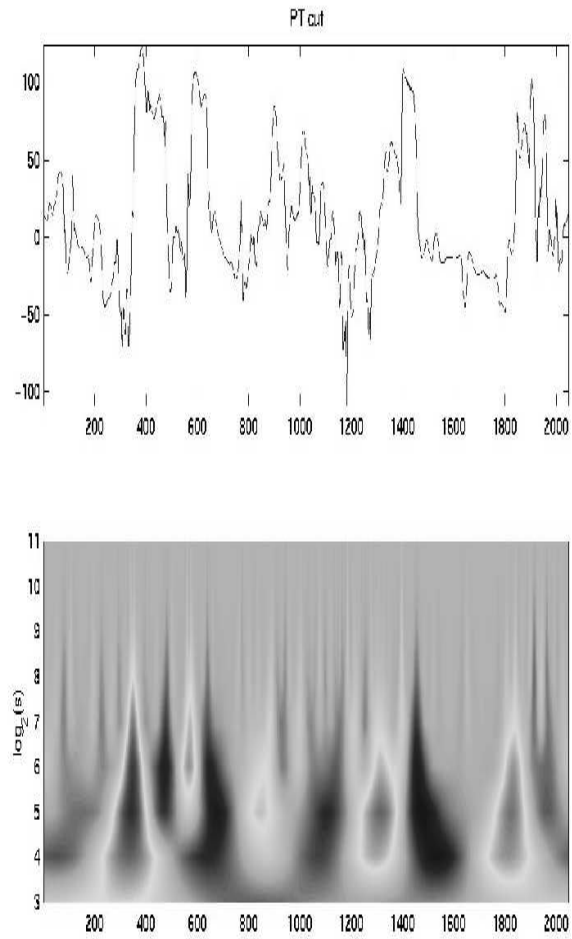}
\caption{Upper Panel : A random 1D cut of the PT field. Lower Panel : Its wavelet transform
(the analysing wavelet is the first derivative of a Gaussian),
the cone like features indicate roughness in the analyzed signal.}
\label{fig:fig4a}
\end{center}
\end{figure}

\begin{figure}
\begin{center}
\epsfxsize=9.0 cm
\epsfysize=14.0 cm
\leavevmode\epsfbox{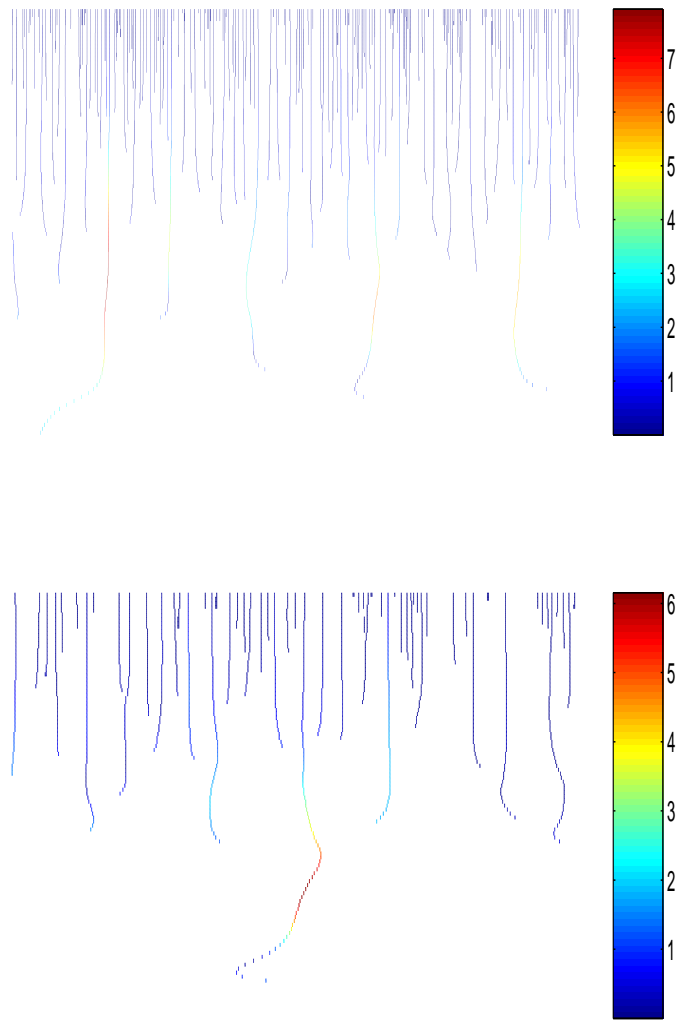}
\caption{Upper Panel : The wavelet transform modulus maxima lines. The lower panel is a zoom into a
particular section of the upper panel, this {\it qualitatively} indicates the sparseness of
the rough spots. A log plot of the magnitude of the maxima line yields the local Holder exponent.}
\label{fig:fig4b}
\end{center}
\end{figure}

By using wavelets whose higher moments vanish one can 
detect sigularities in the higher order derivatives of 
the signal being analyzed \cite{Mallat-IEEE}. 
Our previous
results indicate that the PT field becomes rough, i.e., the first derivatives of the PT
field should be unbounded. Hence,
the particular wavelet we use is the first derivative of a Gaussian 
whose first moment vanishes \cite{Muzy-PRE},i.e., it picks up points where the 
signal becomes rough. 
The wavelet transform of a 1D cut of
the PT field can be seen in Fig. \ref{fig:fig4a}. The cone like features imply the presence
of a rough spot \cite{Mallat-IEEE}. The modulus maxima lines are extracted from this
transformed field and can be seen in Fig. \ref{fig:fig4b}. The value of the local Holder exponent
can be extracted via a log plot of the magnitude of a particular maxima line \cite{Mallat-IEEE}.
In essence, the presence of the cones in the wavelet transform 
indicate roughness in the PT field and
the WTMM lines locate the positions of the rough spots.
However, a closer look (the lower panel of Fig. \ref{fig:fig4b}) suggests, 
{\it qualitatively}, that the rough regions are sparsely distributed (for eg.
comparing with Fig. 8 in Arneodo et. al \cite{Arn-PhyA}).
This goes along with the observation in the last subsection 
that the rough regions were
non space filling. 
In contrast, it has been found \cite{Arn-PhyA},\cite{Muzy-PRL},\cite{Daoudi-book} 
that the local Holder exponents ($h(x)$) associated with a 1D cut of 
the velocity field in 3D turbulence satisfy
$-0.3 \le h(x) \le 0.7$ for almost all $x$ (the peak of the histogram being at $\frac{1}{3}$) 
implying that the velocity field in 3D turbulence is very rough or
has unbounded first derivatives at almost all points. 
\\

\subsection{The Dimension of Level Sets}
Apart from being physically interesting, the level set (iso-$\theta$ contour) dimensions 
provide another link to the roughness of the scalar field.
As the initial condition was a smooth 2D field, initially,
any given level set ($E_{\theta_0}$ for $\theta=\theta_0$) is a non space filling curve, i.e., the level set 
dimension ($D(E_{\theta_0})$) is one. 
In the case of 3D turbulence there exist analytical
estimates of the scalar level set dimensions \cite{Const-PRL}. 
These estimates are
seen to be numerically satisfied by a host of fields (both passive and active) 
in isotropic 3D and Magnetohydrodynamic (MHD) turbulence \cite{Brand-PRA}. 
It must be emphasized that these estimates come
directly from the equations of evolution and are much more powerful than the
phenomenological ideas we have been working with so far. The actual calculation
\cite{Const-PRL},\cite{Const-PRE} is of the area of an isosurface contained in a 
ball of specified size, 
given a Holder condition on the velocity field,

\begin{equation}
        |u(x+y,t) - u(x,t)| \le U_L {(\frac{y}{L})}^{\zeta_u}
\label{10a}
\end{equation}
this area estimate leads to the bound,

\begin{equation}
	D(E_{\theta}) \le 2.5 + \frac{\zeta_u}{2}
\label{10b}
\end{equation}
The
bound in Eq. (\ref{10b}) is expected to be
saturated \cite{Const-PRE} above a certain cutoff scale.
A valid extrapolation 
\cite{Const-Per} for the level sets 
of PT in the SQG system reads,
$D(E_{\theta}) \le 1.5 + (\zeta_1)/2$, where $\zeta_1$ is given by Eq. (\ref{7}) due to the 
equality of the scaling exponents for the velocity field and the PT in the SQG system. 
\\

\begin{figure}
\begin{center}
\epsfxsize=9.0 cm
\epsfysize=14.0 cm
\leavevmode\epsfbox{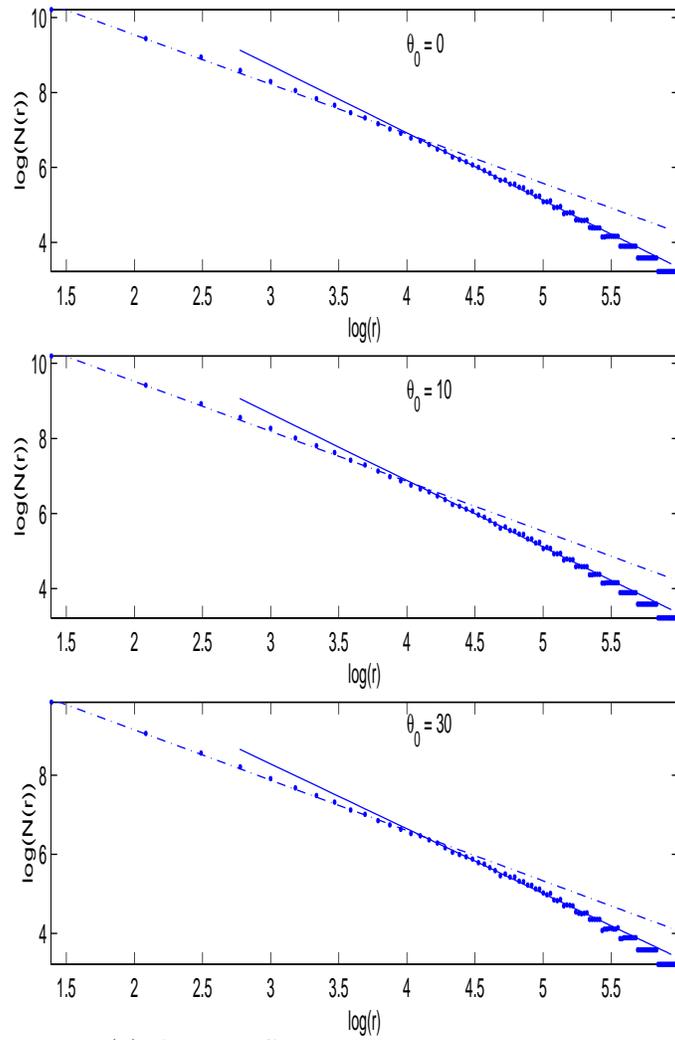}
\caption{$\log(N(r))$ Vs. $\log(r)$ for 3 different level sets. The dashed line is the best fit
for $4 \le r \le 60$, the solid line is the best fit for $64 \le r \le 200$. The box counting
dimensions are $(1.32,1.8), (1.33,1.77) \textrm{and} (1.27,1.64)$ for the small and large $r$
regions for the 3 level sets respectively.}
\label{fig:fig5}
\end{center}
\end{figure}

In passing we mention that the analytical estimates are for the Hausdorff dimension 
whereas
practically we compute the box counting dimension. 
We performed calculations for a variety 
of nominal (i.e., near the mean) level sets and
Fig. \ref{fig:fig5} shows the log-log plots used in the calculation
of the box counting dimension. As is the case for 3D \cite{Brand-PRA},\cite{Const-PRL} there appears to be
a crossover in $D(E_{\theta})$, we find that the even though the dimension undergoes a 
change, the fractal nature seems to persist at smaller scales. For small $r$, $D(E_{\theta}) \approx
1.3$ whereas for larger values of $r$ we find $D(E_{\theta}) \approx 1.74$
(see the caption of Fig. \ref{fig:fig5} for details on the actual values of $r$).
>From our previous
calculations $\zeta_1 = 0.57$,
hence the analytical prediction 
is $D(E_{\theta}) \le 1.785$. Furthermore, as the bound is expected to be saturated above
the cutoff we see that the computed value of $D(E_{\theta})$ for large $r$ is quite close
to the analytical prediction. 
In all, apart from satisfying the analytically prescribed bounds
(and indirectly indicating roughness in the PT field),
the level set dimensions indicate that initially non space filling level sets acquire
a fractal nature in finite time.
\\

\section{Gradient Field Characterization}
The effect of the inferred roughness in the PT field will, as mentioned previously, be
reflected in the singular nature of the gradient fields. In this section we proceed to 
examine a variety of fields which are functions of the PT gradient. The aim is to
see if we can actually detect the expected singularities, and if so, to characterize them.

\subsection{The Dissipation Field}

A physically interesting
function of the gradient field is the PT dissipation, as it is connected to the 
variance of the PT field.
The equation for the dissipation of $\theta$ can be obtained by multiplying Eq. (\ref{1})
by $\theta$ and averaging over the whole domain,

\begin{equation}
        \frac {\partial<\theta^2>}{\partial t} = -2 \nu <(\nabla\theta)^2>
\label{11}
\end{equation}
Here $\nu (\nabla\theta)^2$ is the dissipation field. 
Now consider the quantity $\mu(x,r)$,

\begin{equation}
        {\mu}(x,r) = \frac{1}{r^d} \int_{B(x,r)}{\nu (\nabla\theta)^2}d^dx
\label{12}
\end{equation}
Physically this is the average dissipation in a ball of size $r$ centered at $x$.
Due to the smoothing via integration it is expected that $\mu(x,r)$ will 
be fairly well behaved through most of the domain 
with intermittent bursts of high values concentrated in the regions where the PT field
is rough. The multifractal formalism \cite{Halsey-PRA},\cite{Frisch-book} 
,\cite{Mene-JFM} provides a convenient way
to characterize such "erratic" or singular measures.
The technique \cite{Halsey-PRA} consists of constructing a measure
($\mu$ with suitable normalization) and using its moments to focus on the singularities of the
measure. The domain in which the field is defined is partitioned into disjoint boxes of size
$r$ and it is postulated (see for eg. \cite{Hosokawa-Proc}) that moments of $\mu$ will scale
as,

\begin{equation}
        \sum\mu(x,r)^q \sim r^{\tau_q - qd}
\label{13}
\end{equation}
where the sum goes over all the boxes into which the domain was partitioned. Consider the
set of points where the measure scales $r^{\alpha}$ and denote the dimension of
this set by $f(\alpha)$ (again a probabilistic view is more precise). By similar
considerations as for the $g(\beta)$ spectrum it can be seen that \cite{Frisch-book},

\begin{equation}
        \tau_q = \min_{\alpha}(q\alpha - f(\alpha)) ~ \textrm{and} ~ f(\alpha)=\max_{q}(q\alpha - \tau_q)
\label{14}
\end{equation}
The function $\tau_q$ is further related to the generalized dimensions
\cite{Hent-PhyD} via $D_q=\tau_q/(q-1)$. Practically \cite{Mene-JFM} a log-log plot of the
ensemble average of $\mu(x,r)^q$ for different values of $r$ gives $(D_q-d)(q-1)$. By knowing
$(q,D_q)$ one can use Eq. (\ref{14}) to obtain the $(\alpha,f(\alpha))$ spectrum. 
\\

\begin{figure}
\begin{center}
\epsfxsize=9.0 cm
\epsfysize=9.0 cm
\leavevmode\epsfbox{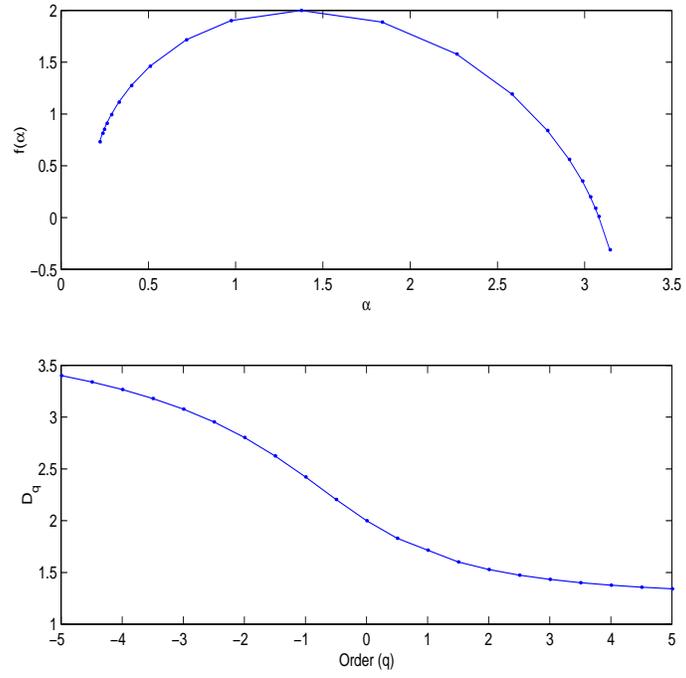}
\caption{The $f(\alpha)$ spectrum and $D_q$ for the dissipation field}
\label{fig:fig6}
\end{center}
\end{figure}

Again, as we expect the scaling of any physical quantity to be restricted to a range of length
scales (say $r_a$ to $r_b$), it is preferable to work in terms of ratios of $\frac{r}{r_b}$ where
$r_b$ is the outer scale, $r_a$ is the inner scale and $r_b \ge r \ge r_a$. In these terms
the generalized dimensions can be expressed as \cite{Mene-JFM},

\begin{equation}
	<{\mu(r)}^q> = C_2(q) ({\mu(r_b)}^q) {(\frac{r}{r_b})}^{(D_q-d)(q-1)}
\label{15}
\end{equation}
where $\mu(r_b)$ is the measure on the outer scale and $C_2(q)$ is order unity for all $q$
(a similar relation would hold if we replaced the $({\mu(r_b)}^q)$ in the right
hand side by
$<{\mu(r)}>^q$ but with different $C_2(q)$'s). 
For $r < r_a$ the dissipation field is assumed to have become smooth via the action
of viscosity. The $f(\alpha)$ spectrum and the generalized dimensions for the 
dissipation field can be seen in Fig. \ref{fig:fig6}. The nontrivial behaviour of the generalized
dimensions demonstrates that the dissipation field is singular (with different singularity
strengths) within these scales. Also, from the calculations we find
that $f(1) < 2$ which acts as a check that the dissipation 
in a finite volume is bounded (as $f(1)$ is the dimension of the 
support of the singular regions) \cite{Sreeni-PRA}. 
\\

\subsection{General Considerations}
The gradient squared nature of the dissipation field 
implied that it was positive definite. In principle one could conceive of 
fields which possess singularities but aren't sign definite. In order to characterize
this possible sign indefiniteness Ott and co-workers studied
\cite{Ott-PRL},\cite{Du-PhyD}
sign singular measures and a related family of exponents called the cancellation
exponents ($\kappa_q$). It is immediately clear that singular nature is a 
prerequisite for the phenomenon of cancellation, the reason being that a nonsingular
field is bounded and we can always add a constant so as to make the field positive
definite and hence eliminate cancellation. Consider a sign indefinite 1D field $\theta(x)$ (which will later
be interpreted as a 1D cut of the PT field) and construct,

\begin{equation}
	{\mu}'(x,r) = (\frac{1}{r}\int_{x}^{x+r}{|\frac{d\theta}{dx'}|dx'}) ~;~
\label{16}
\end{equation}

\begin{equation}
	{\eta}(x,r) = |\frac{1}{r}\int_{x}^{x+r}{(\frac{d\theta}{dx'})dx'}|
\label{17}
\end{equation}
As ${\mu}'$ is defined for the magnitude of the gradient field we can postulate 
(similar to Eq. (\ref{15}) for $\mu$),

\begin{equation}
	<{{\mu}'(r)}^q> = C_3(q) ({{\mu}'(r_d)}^q) {(\frac{r}{r_d})}^{(D'_q-d)(q-1)}
\label{18}
\end{equation}
Here $d=1$ and $D'_q$ are the generalized dimensions associated with 
${\mu}'$. As usual the scaling is restricted to a range of scales ($r_c < r < r_d$) and
$r_c$ is the scale below which ${\mu}'$ appears
smooth. Formally, the entity $\eta$ after 
suitable normalization is a sign singular measure (see Ott et. al. \cite{Ott-PRL} for
a rigorous definition). It was conjectured \cite{Ott-PRL} that the sign singular entity $\eta$ might also
possess scaling properties in analogy with ${\mu}'$, implying, 

\begin{equation}
	<{\eta}(r)^q> = C_4(q) (r)^{-{\kappa}'_q} ~;~ r>r_{cc} 
\label{18b}
\end{equation}
Using Eq. (\ref{18}) we can express Eq. (\ref{18b}) as,

\begin{equation}
<{\eta}(r)^q> \sim <{{\mu}'}(r)^q> (r)^{-\kappa_q} ~;~ r_d > r>r_{cc} \ge r_c 
\label{19}
\end{equation}
Where $\kappa_q = {\kappa}'_q + (D'_q-d)(q-1)$ and $r_{cc}$ is the 
lower oscillatory scale below which the derivative does not
oscillate. 
We prefer to call $\kappa_q$ from Eq. (\ref{19})
the cancellation exponents as they directly reflect the difference in 
the scaling properties of $\frac{d\theta}{dx}$ and $|\frac{d\theta}{dx}|$.
\\

Apriori there is no justification in assuming that the scale at which cancellation
ceases ($r_{cc}$) is the same as the scale at which ${\mu}'$ becomes smooth ($r_c$).
Consider the example where the derivative is a discrete signal 
composed of a train of delta functions (zero elsewhere)
where the minimum separation between the delta functions is $l$. Furthermore 
let us assign the sign of the delta functions in a random fashion. In this case
the $r_{cc} = l$, in fact, $\kappa_1 = 0.5$ due to the random distribution
of the signs. But 
as the delta functions are supported at points we have $r_c \rightarrow 0$.
As an aside we point out that if there
is a maximum scale of separation between the delta functions (say $L$) then at
scales greater than $L$ we will see $D'_q = d$ for ${\mu}'$. The reason for
pointing this out is to give a feel for fields that exhibit scaling, on one
hand smooth fields have $D'_q = d$ whereas on the other extreme random fields with 
small correlations also have $D'_q = d$ (at scales larger than their correlation lengths). 
Fields with nontrivial scaling over a significant 
range are in effect
random but with large correlation lengths. The reader is referred to Marshak et. al. 
\cite{Marshak-PRE} for
a detailed examination of multiplicative processes and the resulting characterization
by structure functions and generalized dimensions.
\\

In order to get a unified picture of scaling in both the gradient and the field itself
there have been attempts (for eg. \cite{Vain-PRE},\cite{Bert-PRE}) to link $\zeta_q$, $\kappa_q$ 
and $D'_q$ to each other. The view that seems to have emerged is that there exist 
simple relations linking the various exponents and that these relations are valid under very
general conditions.
We present an alternate derivation of 
some of these relations which
makes the implicit assumptions explicit. 
Proceeding from Eq. (\ref{19}), assuming that
$\frac{d\theta}{dx}$ has integrable singularities we obtain,

\begin{equation}
	<|\theta(x+r) - \theta(x)|^q> = <{{\mu}'}(r)^q> r^{q-\kappa_q}
\label{20}
\end{equation}
Substituting from Eq. (\ref{18}) and on comparing with Eq. (\ref{7}) we get,

\begin{equation}
	\zeta_q = (D'_q-d)(q-1) + q - \kappa_q
\label{21}
\end{equation}
Writing Eq. (\ref{20}) for $q=1$ we have,

\begin{equation}
	<|\theta(x+r) - \theta(x)|> = <{{\mu}'}(r)> r^{1-\kappa_1}
\label{22}
\end{equation}
which yields $\kappa_1 + \zeta_1 = 1$. Now if we make a strong assumption regarding
the uniform nature of the cancellation, it is possible to claim that Eq. (\ref{22}) holds 
not only in average but on every interval, ie,

\begin{equation}
	|\theta(x+r) - \theta(x)| = {{\mu}'}(r) r^{1-\kappa_1}
\label{23}
\end{equation}
Raising this to the $q^{th}$ power and performing an ensemble average yields,

\begin{equation}
	<|\theta(x+r) - \theta(x)|^q> = <{{\mu}'}(r)^q> r^{q(1-\kappa_1)}
\label{24}
\end{equation}
which implies, 

\begin{equation}
	\zeta_q = (D'_q-d)(q-1) + q (1- \kappa_1)
\label{25}
\end{equation}
The implications of Eq. (\ref{25}) are quite severe in that 
it shows $\zeta_q$ to be dependent on $D'_q$ and the knowledge of only the first
cancellation exponent allows the derivation of one from the other. In general
the scaling exponents of $\theta$ and the generalized dimensions of ${\mu}'$
provide exclusive information.  It is only in the presence of integrable 
singularities that one can link the two via Eq. (\ref{21}), furthermore the stronger
relation (Eq. (\ref{25})) is valid under the 
added assumption of uniform cancellation. 
\\

\subsection{The Gradient of the PT and its absolute value}

\begin{figure}
\begin{center}
\epsfxsize=9.0 cm
\epsfysize=14.0 cm
\leavevmode\epsfbox{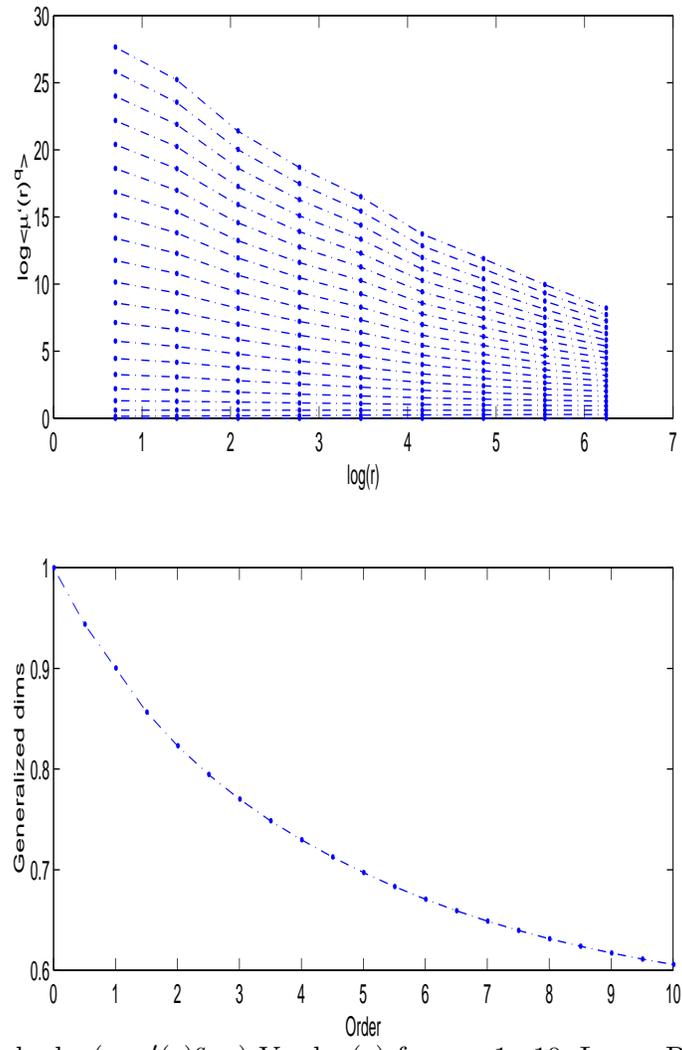}
\caption{Upper Panel : $\textrm{log}(<{\mu}'(r)^q>)$ Vs. $\textrm{log}(r)$ for $q=1:10$, Lower Panel : The generalized
dimensions for ${\mu}'$}
\label{fig:fig7}
\end{center}
\end{figure}

\begin{figure}
\begin{center}
\epsfxsize=9.0 cm
\epsfysize=9.0 cm
\leavevmode\epsfbox{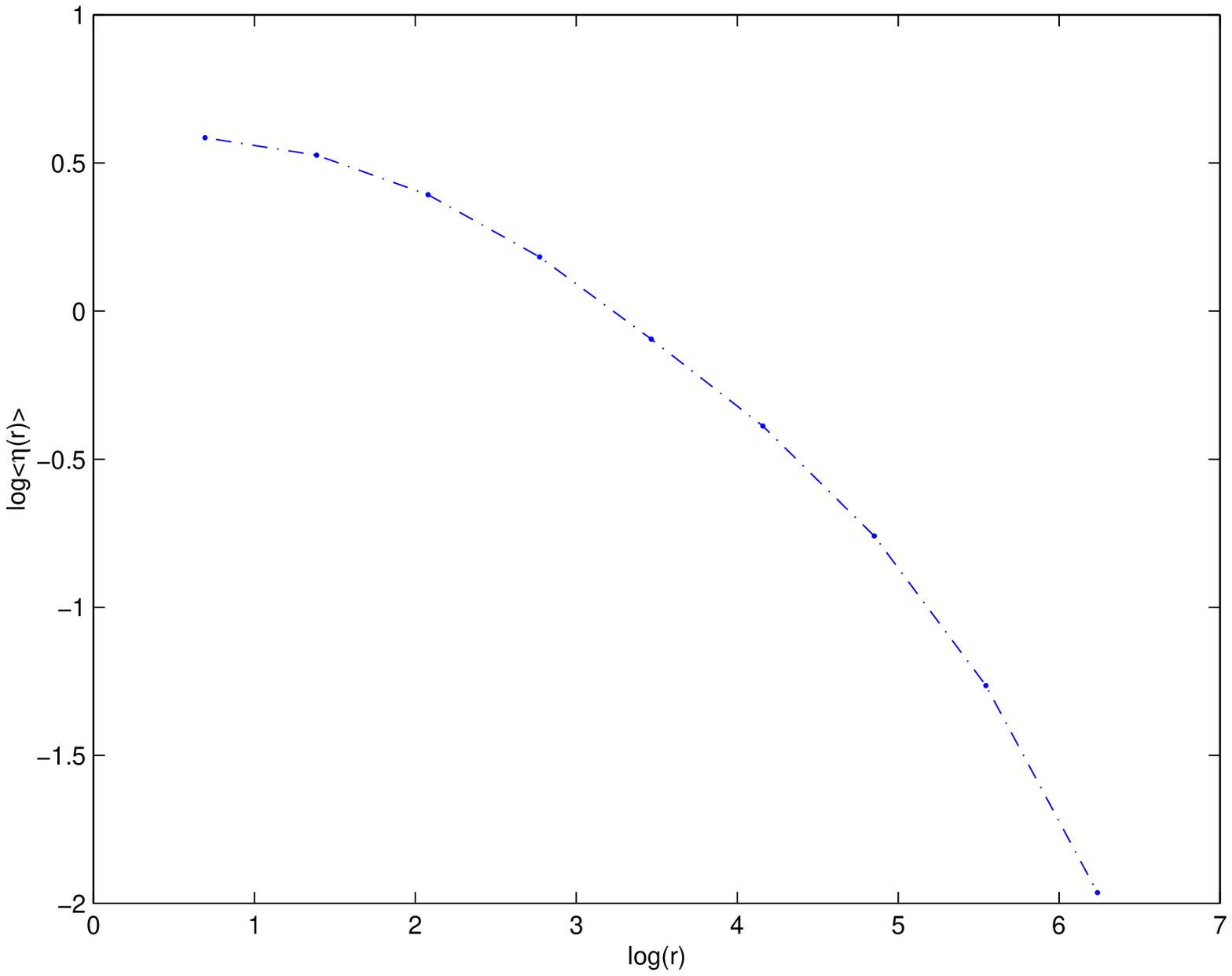}
\caption{$\textrm{log}(<{\eta}(r)>)$ Vs. $\textrm{log}(r)$}
\label{fig:fig8}
\end{center}
\end{figure}

We proceed to check if scaling is 
observed (as postulated) for the PT field gradient and its absolute value and 
whether one can extract the aforementioned
exponents. In the upper panel of Fig. \ref{fig:fig7} we show the log-log plots of 
$<{\mu}'(r)^q>$ Vs. $r$ for different $q$. The scaling relations certainly 
appear to hold true (which was expected as they held for the dissipation field). The
generalized dimensions for ${\mu}'$ can be seen the lower panel of Fig. \ref{fig:fig7}.
On the other hand the scaling for $<{\eta}(r)>$ seen in Fig. \ref{fig:fig8}
fails to exhibit a 
power law in $r$. Hence there is no 
meaningful way of extracting the cancellation exponents as they have been defined.
Unfortunately this implies that the relations derived in the previous section
(Eqs. (\ref{21}) and (\ref{25})) cannot be used in this situation. In spite of this, we can see
that as $r$ decreases the tangent to $\textrm{log}(<{\eta}(r)>)$ has a smaller slope
which is consitent with the existence of a oscillatory cutoff at small scales. 
\\

\section{Conclusion and Discussion}
In summary, we have found that the PT field in the SQG equations appears to become rough 
within a specified range of scales. Moreover, not only is there a heirarchy in the degree of 
roughness, the roughness is distributed sparsely in a qualitative sense.
These conclusions are based on a combination of factors, namely, 
the algebraic power spectrum, anomalous scaling in the structure functions, a nontrivial
$(\beta,g(\beta))$ spectrum, the nature of the WTMM map and the wrinkling of the PT level sets.
The roughness in the PT field is expected to have an adverse effect on functions of the gradient 
field. This expectation is bourne out in the multifractal nature of the dissipation
field. Also, the singular nature of the gradient field in combination with its 
sign indefinetness led us to examine a sign singular measure based upon the 
gradient field. The failure to observe scaling in the sign singular measure serves,
in our opinion, as a reminder that most scaling arguments are postulated at a 
phenomenological level and the underlying basis of why scaling is observed in 
the first place is a fairly subtle and unsettled issue. Similarly the simple 
derivation of the relation between the variety of scaling exponents makes explicit
some of the assumptions that are required for the validity of similar relations proposed
in earlier studies.
\\

Regarding the more general question we posed in the very beginning of this paper, i.e., where
does the SQG active scalar stand with respect to both 2D and 3D turbulence, we have the 
following comments. With respect to the vorticity in the 2D Euler equations the corresponding
quantity in the SQG system is the PT. The coherent structure/fine background nature of the 
vorticity field \cite{Benzi-EPL},\cite{Benzi-Math},\cite{Mizu-Jap} carries over qualitatively
to the PT field. 
The vorticity 
structure functions which showed normal scaling in the 2D Euler system \cite{Benzi-EPL}
cross over to anomalous scaling in the SQG system. 
Our conjecture is that the stronger
local interactions in
the SQG system are responsible for this anomalous scaling.
In both these systems the vorticity
and PT respectively are conserved quantities and hence any singularities one might
experience are actually in the gradient fields, as is seen in the multifractal nature
of the enstrophy dissipation in the Euler system \cite{Benzi-EPL} and the PT dissipation
in the SQG system.
\\

In the 3D Euler case, the PT from the SQG equations is analogous to the velocity field and 
$\vec{{\nabla}^{\perp}\theta}$ from SQG corresponds to the vorticity of the 3D Euler
equations. The roughness
of the PT field is similar to the postulated roughness of the velocity
field in the inertial range, a subtle difference being that the roughness in the PT 
appears to be 
sparse whereas indications are that the roughness in the 3D velocity field is present
almost everywhere \cite{Arn-PhyA},\cite{Muzy-PRL},\cite{Daoudi-book}. We re-emphasize that
in both these cases the roughness is restricted to a range of scales, i.e., no claim
is made for an actual singularity in the corresponding gradients. Similarly 
the anomalous scaling of the PT follows that of the velocity field in 3D but again it
is not as strong as in the 3D case. The general theory
developed \cite{Const-PRL} for the deformation of scalar level sets in the 3D case 
is seen to carry over to the SQG equations. In essence the SQG equations follow the 
3D Euler equations but in a somewhat weaker sense. This "weakness" is clearly manifested in the
behaviour of the gradients. In the 3D Euler equations a sign singular measure
constructed from the vorticity field
shows good scaling properties and a cancellation exponent can be meaningfully
extracted \cite{Ott-PRL}, whereas in the SQG equations a similarly constructed 
entity lacks scaling.
\\

The results presented in this paper have been for the most part diagnostic, in that
they characterize the nature of the roughness of the potential temperature field
in SQG dynamics. Although we have exhibited anomalous scaling of the potential
temperature fluctuations, we do not have a theory accounting for the observed
form of $\zeta_q$.  Arriving at such a theory will be a major challenge for
future work.  Our results point efforts in the direction of considering the
dichotomy between smooth fields within large organized vortices, and a rather
sparse set at the boundaries of and between vortices which exhibits a greater
degree of roughness.  The diagnostic results also suggest a means for distinguishing
between SQG and Euler dynamics in Nature, in cases where only a tracer field
can be observed, as in the gas giant planets (notably Saturn, Jupiter and Neptune,
which exibit a rich variety of turbulent patterns).  SQG dynamics should yield
anomalous scaling corresponding to sparse roughness, whereas Euler dynamics should
yield normal scaling.
\\

\acknowledgments
We thank Prof. F. Cattaneo for performing the actual numerical simulations of
the SQG system. One of the authors (J.S.) would like to thank Prof. N. Nakamura
for advice regarding the manuscript and Prof. P. Constantin for clarifying
certain points. Wavelab 805 (http://www-stat.stanford.edu/~wavelab/) was used for 
certain parts of the wavelet analysis. This project is supported by the ASCI Flash Center at the
University of Chicago under DOE contract B341495


\clearpage

\begin{references}

\bibitem{Pedlosky} J. Pedlosky. 
{\em Geophysical Fluid Dynamics},
Springer Verlag, Chapter 6, 1979.

\bibitem{Held-JFM} I. Held, R. Pierrehumbert, S. Garner and K. Swanson. 
{\em Surface quasi-geostrophic dynamics},
Journal of Fluid Mechanics, {\bf 282}, 1, 1995.

\bibitem{RTP-CSF} R. Pierrehumbert, I. Held and K. Swanson. 
{\em Spectra of Local and Nonlocal Two-dimensional Turbulence},
Chaos, Solitons and Fractals, {\bf 4}, 6, 1111, 1995.

\bibitem{Const-NL} P. Constantin, A. Majda and E. Tabak. 
{\em Formation of strong fronts in the 2D quasigeostrophic thermal active scalar}, 
Nonlinearity, {\bf 7}, 1495, 1994.

\bibitem{Majda-P} A. Majda and E. Tabak. 
{\em A two-dimensional model for
for quasigeostrophic flow: comparison with two-dimensional Euler flow},
Physica D, {\bf 98}, 515, 1996.

\bibitem{Sch} N. Schorghofer. {\em Energy spectra of steady two-dimensional turbulent flows},
Physical Review E, {\bf 61}, 6572, 2000.

\bibitem{Const-PRE1} P. Constantin, Q. Nie and N. Schorghofer.
{\em Front formation in an active scalar equation},
Physical Review E, {\bf 60}, 3, 2858, 1999.

\bibitem{Okhitani-PHF} K. Okhitani and M. Yamada. 
{\em Inviscid and inviscid-limit behaviour of a surface quasigeostrophic flow},
Physics of Fluids, {\bf 9}, 4, 876, 1997.

\bibitem{Cor} D. Cordoba and C. Fefferman. {\em Behaviour of several 
two-dimensional fluid equations in singular scenarios},
Proc. Nat. Acad. Sci. USA, {\bf 98}, 4311, 2001.

\bibitem{Benzi-EPL} R. Benzi and R. Scardovelli. 
{\em Intermittency of Two-Dimensional Decaying Turbulence},
Europhys. Lett., {\bf 29}, 5, 371, 1995.

\bibitem{Benzi-Math} R. Benzi, G. Paladin, S. Patarnello, P. Santangello and A. Vulpiani. 
{\em Intermittency and coherent structures in two-dimensional turbulence},
J. Phys. A : Math. Gen., {\bf 19}, 3371, 1986.

\bibitem{Mizu-Jap} H. Mizutani and T. Nakano. 
{\em Multifractal Analysis of Simulated Two-Dimensional Turbulence},
Journal of the Physical Society of Japan, {\bf 58}, 5, 1595, 1989.

\bibitem{Vain-PRE}S. Vainshtein, K. Sreenivasan, R. Pierrehumbert, V. Kashyap and A. Juneja.
 {\em Scaling exponents for turbulence and other random processes and their
relationships with multifractal structure},
Physical Review E, {\bf 50}, 3, 1823, 1994.

\bibitem{Rose-JPP} H. Rose and P. Sulem. 
{\em Fully Developed Turbulence and Statistical Mechanics},
J. Phys. Paris, {\bf 47}, 441, 1978.

\bibitem{Benzi-PRA} R. Benzi, G. Paladin and A. Vulpiani. 
{\em Power spectra in two-dimensional turbulence},
Physical Review A, {\bf 42}, 6, 3654, 1990.

\bibitem{Frisch-book} U. Frisch.
{\em Turbulence},
Cambridge Press, 1995.

\bibitem{Bert-PRE} A. Bertozzi and A. Chhabra. {\em 
Cancellation exponents and fractal scaling},
Physical Review E, {\bf 49}, 5, 4716, 1994.

\bibitem{Frisch-Proc} U. Frisch. 
{\em From global scaling, a la Kolmogorov, to local multifractal scaling in fully
developed turbulence},
Proc. R. Soc. Lond. A, {\bf 434}, 89, 1991.

\bibitem{Muzy-PRE} J. Muzy, E. Bacry and A. Arneodo. {\em
Multifractal formalism for fractal signals: The structure-function approach versus
the wavelet-transform modulus-maxima method},
Physics Review E, {\bf 47}, 2, 875, 1993.

\bibitem{Arn-PhyA} A. Arneodo, E. Bacry and J. Muzy. {\em
The thermodynamics of fractals revisited with wavelets},
Physica A, {\bf 213}, 232, 1995.

\bibitem{Muzy-PRL} J. Muzy, E. Bacry and A. Arneodo. {\em
Wavelets and Multifractal Formalism for Singular Signals: Application to
Turbulence Data},
Physical Review Letters, {\bf 67}, 25, 3515, 1991.

\bibitem{Mallat-IEEE} S. Mallat and W. Hwang. {\em 
Singularity Detection and Processing with Wavelets},
IEEE Trans. on Information Theory, {\bf 38}, 2, 617, 1992.

\bibitem{Daoudi-book} K. Daoudi. {\em
A New Approach for Multifractal Analysis of Turlbulence Signals},
Fractals and Beyond, M. Novak
Ed., World Scientific, 91, 1998.

\bibitem{Const-PRL} P. Constantin, I. Proccacia and K. Sreenivasan. {\em 
Fractal Geometry of Isoscalar Surfaces in Turbulence: theory and experiments},
Physical Review Letters, {\bf 67}, 13, 1739, 1991.

\bibitem{Brand-PRA} A. Brandenburg, I. Proccacia, D. Segel and A. Vincent. {\em 
Fractal level sets and multifractal fields in direct simulation of turbulence},
Physical Review A, {\bf 46}, 8, 4819, 1992.

\bibitem{Const-PRE} P. Constantin and I. Proccacia. {\em 
Scaling in fluid turbulence : A geometric theory},
Physical Review E, {\bf 47}, 5, 3307, 1993.

\bibitem{Const-Per} P. Constantin. Personal communication.

\bibitem{Halsey-PRA} T. Halsey, M. Jensen, L. Kadanoff, I. Procacia and B. Shraiman. {\em 
Fractal measures and their singularities: the characterization of strange sets},
Physics Review A, {\bf 33}, 1141, 1986.

\bibitem{Mene-JFM} C. Meneveau and K. Sreenivasan. {\em 
The multifractal nature of turbulent energy dissipation},
Journal of Fluid Mechanics, {\bf 224}, 429, 1991.

\bibitem{Hosokawa-Proc} I. Hosokawa. {\em 
Theory of scale-similar intermittant measures},
Proc. Roy. Soc. London A, {\bf 453}, 691, 1997.

\bibitem{Hent-PhyD} H. Hentschel and I. Procaccia. {\em 
The infinite number of generalized dimensions of fractals and strange attractors},
Physica D, {\bf 8}, 435, 1983.

\bibitem{Sreeni-PRA} K. Sreenivasan and C. Meneveau. {\em 
Singularities of the equations of fluid motion},
Physical Review A, {\bf 38}, 12, 6287, 1988.

\bibitem{Ott-PRL}E. Ott, Y. Du, K. Sreenivasan, A. Juneja and A. Suri. {\em 
Sign-Singular Measures: Fast Magnetic Dynamos, and High-Reynolds-Number Fluid Turbulence},
Physical Review Letters, {\bf 69}, 18, 2654, 1992.

\bibitem{Du-PhyD} Y. Du, T. Tel and E. Ott. {\em 
Characterization of sign-singular measures},
Physica D, {\bf 76}, 168, 1994.

\bibitem{Marshak-PRE} A. Marshak, A. Davis, R. Cahalan and W. Wiscombe. {\em 
Bounded cascade models as nonstationary multifractals},
Physics Review E, {\bf 49}, 1, 55, 1994.

\end{references}
\end{document}